\documentclass[aps,twocolumn,showpacs,amsmath,amssymb,prb,reprint]{revtex4-1}
\usepackage{graphicx}%
\usepackage{dcolumn}
\usepackage{bm}
\usepackage{braket}
\begin{document}
\title{Analytical formulation for the modulation of the time-resolved dynamical Franz-Keldysh effect by the electron excitation in dielectrics}
\author{T. Otobe}
\affiliation{Kansai Photon Science Institute, National Institutes for Quantum and Radiological Science and Technology (QST), Kyoto 619-0215, Japan}
\begin{abstract}
Analytical formulation of sub-cycle modulation (SCM) of dielectrics including electron excitation is presented.
The SCM is sensitive to not only the time-resolved dynamical Franz-Keldysh effect (Tr-DFKE)
[T. Otobe, {\it et al.}, Phys. Rev. B {\bf 93}, 045124 (2016)], which is the nonlinear response without the electron excitation,
but also the excited electrons.
The excited electrons enhance the modulation with even-harmonics of pump laser frequency, 
and generate the odd-harmonics components.
The new aspect of SCM is a consequence of i) the interference between the electrons excited by the pump laser and those excited by the probe pulse laser
and ii) oscillation of the generated wave packed by the pump laser. 
When the probe- and pump-pulse polarizations are parallel, the enhancement of the even harmonics  and the generation of the 
odd harmonics modulation appear.
However, if the polarizations are orthogonal, the effect arising from the electron excitations becomes weak.
By comparing the parabolic and cosine band models, I found that the electrons under the intense laser field move as quasi-free particles.
\end{abstract}
\maketitle
\section{introduction}
Since the beginning of this century, the intense research into 
attosecond light pulses has launched a new era in ultrafast material science \cite{atto01,Chen12}.
In particular, the technique of  attosecond transient absorption spectroscopy (ATAS) is now being used to observe electron dynamics in solids 
\cite{Cavalieri07,Krausz13,Schultze13,Schultze14,Neppl15,Sommer16,Mashiko16,Lucchini16,Michael17}.

There are  two notable aspects of ATAS, specifically, 
the sub-cycle modulation (SCM)  of the optical properties of solids using an intense pulse laser field \cite{Schultze13,Novelli13,Pati15,Mashiko16,Lucchini16,Uchida16}
and the saturated absorption~\cite{Schultze14,Michael17}.
Sub-cycle modulation is a crucial in so-called peta-Hertz engineering applications \cite{Krausz13,Schultze13,Mashiko16}.  
In previous works, the sub-cycle oscillations of the optical properties evident in 
the time-resolved dynamical Franz-Keldysh effect (Tr-DFKE) was reported \cite{Lucchini16,Franz58,Keldysh58,Jauho96,Nordstorm98}.
This effect manifests as the response of many dressed states at probe time in the absence of electron excitations \cite{otobe16,otobe16-2,otobe16-3,Mizumoto06}.
A similar effect was reported by Uchida {\it et al.} from the excitonic states in GaAs quantum wells~\cite{Uchida16}.
In regard to saturated absorption, this provides detail of  the excited electron-hole pairs \cite{Schultze14,Michael17}.

In this paper, we address analytically the question of how the Tr-DFKE is modulated by electron excitations.
For symmetric materials, the Tr-DFKE results in the ultrafast oscillations of their optical properties 
 with even-order harmonics of the pump laser frequency \cite{Uchida16}.
We found that i) the interference between the electrons excited by the pump laser and those excited by the probe pulse 
 generates the odd-order harmonics  and ii) it enhances the even-order harmonics components  
 if the pump and probe laser polarizations are in parallel.
 If the  polarizations are orthogonal, the modulations in optical properties are small.
  We also examine the parabolic and cosine band models to resolve the band structure.
  We found that the former shows good qualitative agreement with previous first-principles calculation \cite{otobe16}, indicating 
  that the wave function corresponds to an accelerated quasi-free particle under an intense laser field.

In the following two section, we derive an analytical expression for the SCM electron excitations using a parabolic two-band model, 
and present numerical results of the SCM for diamond and laser parameter dependences.
We finish with a summary.

\section{Formulation}
Three steps are used to derive the analytical formulas.
The first  considers a spatially periodic system:
\begin{equation}
\varepsilon_{n,\vec{k}}^G u_{n,\vec{k}}^G(\vec{r})=\left[\frac{1}{2}\left(\vec{p}+\vec{k}\right)^2+V(\vec{r})\right]u_{n,\vec{k}}^G(\vec{r}),
\end{equation}
where $\vec{k}$ is the Bloch wave vector and $n$ is the band index. 
We shall assume a simple two-band system $n=c$, $v$, where $c$ ($v$) signifies the conduction (valence) band.

The time-dependent Schr\"odinger equation (TDSE) governing the form of the wave function $u_{n,\vec{k}}(t)$ for a system subject to 
a pump laser field $\vec{A}(t)$ is given as
\begin{equation}
i \frac{\partial u_{n,\vec{k}}(\vec{r},t)}{\partial t}=H(t)u_{n,\vec{k}}(\vec{r},t),
\end{equation}
with time-dependent Hamiltonian
\begin{equation}
H(t)=\left[\frac{1}{2}\left(\vec{p}+\vec{k}+\frac{e}{ c}\vec{A}(t)\right)^2+V(\vec{r})\right].
\end{equation}
The Houston function \cite{Houston} can be used to explore the from of $u_{n,\vec{k}}(t)$ \cite{otobe16,otobe16-2} is written
\begin{equation}
w_{n,\vec{k}}(\vec{r},t)=u_{n,\vec{k}+\frac{e}{c}\vec{A}(t)}^G(\vec{r})\exp\left[-i\int^t dt' \varepsilon_{n,\vec{k}}^G(t') \right]
\end{equation}
where $\varepsilon_{n,\vec{k}}^G(t)=\varepsilon_{n,\vec{k}+\frac{e}{c}\vec{A}(t)}^G$.

The wave function of the valence band (VB) $u_{v,\vec{k}}(t)$ includes the reversible and irreversible transition to the conduction band (CB), 
 comprising wave functions$w_{c,\vec{k}}(\vec{r},t)$\cite{Sommer16}. 
The excitation from VB to CB can be expressed using the Houston functions and complex coefficient of $w_{c,\vec{k}}(\vec{r},t)$.
The time-dependent wave function of the VB can then be expanded as
\begin{equation}
u_{v,\vec{k}}(t)=w_{v,\vec{k}}(\vec{r},t)+C^{\vec{k}}_{vc}(t)w_{c,\vec{k}}(\vec{r},t),
\end{equation}
where the coefficient is given by: 
\begin{eqnarray}
\label{eq:Coe_C}
C^{\vec{k}}_{vc}(t)
&=&-e\int^t dt' \vec{E}(t')\cdot\frac{\vec{P}^{\vec{k}+\frac{e}{c}\vec{A}(t')}}{\varepsilon^G_{c,\vec{k}}(t')-\varepsilon^G_{v,\vec{k}}(t') }e^{i S(t')}
\end{eqnarray}
with
\begin{equation}
S(t)=\int^t dt'\left(\varepsilon_{c,\vec{k}}^G(t') -\varepsilon_{v,\vec{k}}^G(t')\right)
\end{equation}
and
\begin{equation}
\vec{P}^{\vec{k}+\frac{e}{c}\vec{A}(t)}=\left<u_{c,\vec{k}}^{G}\Big|\vec{p}\Big|u_{v,\vec{k}}^G\right>\Big|_{\vec{k}+\frac{e}{c}\vec{A}(t)}.
\end{equation}

In the next step, the TDSE for the VB $\tilde{u}_{v,\vec{k}}(t)$ subject to pump and a weak probe laser $\vec{A}_p(t)$ can be written 
\begin{equation}
i\frac{\partial \tilde{u}_{v,\vec{k}}(\vec{r},t)}{\partial t}=\left[H(t)+\delta H(t)\right]\tilde{u}_{v,\vec{k}}(\vec{r},t),
\end{equation}
where $\delta H(t)$ is treated as a perturbative term of the Hamiltonian,
\begin{equation}
\delta H(t)\approx\frac{e}{c}\left(\vec{p}+\vec{k}+\frac{e}{c}\vec{A}(t)\right)\cdot\vec{A}_p(t).
\end{equation}
The probe pulse is applied to the system at a specified time, $T_p$.
If the photo-emission induced by the probe pulse is negligible, the process we need to consider is the electron excitation from the VB to CB.
The new time-dependent wave function $\tilde{u}_{v,\vec{k}}(\vec{r},t)$ as
\begin{equation}
\tilde{u}_{v,\vec{k}}(\vec{r},t)=u_{v,\vec{k}}(\vec{r},t)+D^{\vec{k}}(t)w_{c,\vec{k}}(\vec{r},t),
\end{equation}
where $D^{\vec{k}}(t)$ is a first-order coefficient of expansion given by:
\begin{eqnarray}
\label{eq:Coe_D}
D^{\vec{k}}(t)&\approx&-\frac{ie}{ c}\int dt' \vec{P}^{\vec{k}+\frac{e}{c}\vec{A}(t')}\cdot\vec{A}_p(t')
e^{ i S(t')}\nonumber\\
&-&\frac{ie}{ c}\int dt' \Big[C^{\vec{k}}(t')\vec{P}^{\vec{k}+\frac{e}{c}\vec{A}(t')}_{cc}\cdot\vec{A}_p(t') \nonumber\\
&+&C^{\vec{k}}(t')\vec{A}_p(t')\cdot\left(\vec{k}+\frac{e}{ c}\vec{A}(t')\right)\Big\}\Big].
\end{eqnarray}
where 
\begin{equation}
\label{Pcc}
\vec{P}^{\vec{k}+\frac{e}{c}\vec{A}(t)}_{cc}=\left<u_{c,\vec{k}}^G\Big| \vec{p}\Big| u_{c,\vec{k}}^G\right>\Big|_{\vec{k}+\frac{e}{c}\vec{A}(t)}
=\frac{\partial \varepsilon_{c,\vec{k}}(t)}{\partial \vec{k}}\Big|_{\vec{k}+\frac{e}{c}\vec{A}(t)}.
\end{equation}

The electron current $\vec{J}(t)$ induced by the pump- and probe-laser pulses is given by:
\begin{eqnarray}
\label{eq:Cur}
\vec{J}(t)
&=&-\frac{e}{V_{cell}}\sum_{\vec{k}}\Re\left<\tilde{u}_{v,\vec{k}}\left|\vec{p}+\vec{k}+\frac{e}{c}\left(\vec{A}(t)+\vec{A}_p(t)\right)\right|\tilde{u}_{v,\vec{k}}\right>\nonumber\\
&\approx&\vec{J}_P(t)  -\frac{e^2}{c}\vec{A}_p(t)N_e
-\frac{2e}{V_{cell}} \sum_{\vec{k}}\Re\Bigg[D^{\vec{k}*}_{vc}\vec{P}^{\vec{k}+\frac{e}{c}\vec{A}(t)}e^{iS(t)} \nonumber\\
&+&D^{\vec{k}*}C^{\vec{k}}\vec{P}_{cc}^{\vec{k}+\frac{e}{c}\vec{A}(t)}+D^{\vec{k}*}C^{\vec{k}}  \left(\vec{k}+\frac{e}{c}\vec{A}(t)\right)\Bigg],
\end{eqnarray}
where $\vec{J}_P$ is the current induced by only the pump laser, $N_e$ is the electron density of the system, and $V_{cell}$ is the cell volume.
Note that $\vec{J}_P(t)$ contains the generated high-harmonics~\cite{Ghimire11}. 
The term $D^{\vec{k}*}_{vc}\vec{P}^{\vec{k}}e^{iS(t)}$  includes the dynamical Franz-Keldysh effect, $\int dt' \vec{P}^{\vec{k}*}\cdot\vec{A}_p(t')\vec{P}^{\vec{k}}e^{i(S(t)-S(t'))}$,
 and the response of excited state, $\int^t dt' \left[C^{\vec{k}*}_{vc}(t')\vec{A}_p(t') \cdot\left(\vec{k}+\frac{e}{c}\vec{A}(t')\right)\right]\vec{P}^{\vec{k}}e^{iS(t)}$.  
The last term of Eq.~(\ref{eq:Cur}) containing $D^{\vec{k}*}_{vc}C^{\vec{k}}_{vc}$  indicates the interference between the electrons excited by the pump laser and those excited by the probe laser.

The observed conductivity $\sigma(\omega)$ induced by the probe light is given by the relationship,
$\sigma(\omega)=(\tilde{J}(\omega)-\tilde{J}_P(\omega))/\tilde{E_p}(\omega)$,
whrere $\tilde{J}$ and $\tilde{J}_P$ are the Fourier transforms of $J$ and $J_P$, respectively, and 
$\tilde{E}_p$ is the electric field of the probe laser.
The real-part of the conductivity $\Re\sigma(\omega)$ corresponds to the photo-absorption, and
 $\sigma(\omega)$ contains the state-specific current components of real materials.
The $\sigma(\omega)$ has three contributing terms, $\sigma(\omega)=\sigma_0(\omega)+\sigma_{DFKE}(\omega)+\sigma_{ex}(\omega)$, where
$\sigma_0(\omega)$ is the conductivity without the pump laser, 
$\sigma_{DFKE}(\omega)$ is the Tr-DFKE, and $\sigma_{ex}(\omega)$ is the new term related to the electron excitation.

\section{Application for diamond}
\subsection{Parabolic band}
To simplify the calculation, the band structure is defined as a parabolic two-band system $\varepsilon_{c}=B_g+k^2/2m_{c}$ and 
$\varepsilon_{v}=-k^2/2m_{v}$ given band gap $B_g$.
Then the $\vec{P}_{cc}^{\vec{k}+\frac{e}{c}\vec{A}(t)}$ defined by the Eq.~(\ref{Pcc}) is given as
\begin{equation}
\vec{P}^{\vec{k}+\frac{e}{c}\vec{A}(t)}_{cc}=\frac{\vec{k}+\frac{e}{c}\vec{A}(t)}{m_c}.
\end{equation}
The coefficients $C^{\vec{k}}(t)$ and $D^{\vec{k}}(t)$ can be written down as
\begin{eqnarray}
\label{eq:Coe_C_para}
C^{\vec{k}}(t)=-\int^t dt'\vec{E}(t')\cdot\frac{\vec{P}^{\vec{k}+\frac{e}{c}\vec{A}(t')} e^{iS(t')}}{B_g+\frac{\left(\vec{k}+\frac{e}{c}\vec{A}(t')\right)^2}{2\mu}},
\end{eqnarray}
and
\begin{eqnarray}
\label{eq:Coe_D_para}
D^{\vec{k}}(t)&=&
-\frac{ie}{c}\int^t dt'\vec{P}^{\vec{k}}\cdot\vec{A}_p(t')
e^{ i S(t')}\nonumber\\
&-&\frac{ie}{c} \left(\frac{ 1}{ m_c}+1 \right) \nonumber\\
&\times&\int^t dt' \left[C^{\vec{k}}(t')\vec{A}_p(t') \cdot\left(\vec{k}+\frac{e}{c}\vec{A}(t')\right)
\right],
\end{eqnarray}
respectively. Here $\mu$ is the reduced mass of the electron-hole state.
The electron current $\vec{J}(t)$ induced by the pump and probe-laser pulses is also given by
\begin{eqnarray}
\label{eq:Cur_para}
\vec{J}(t)
&=&\vec{J}_P(t)  -\frac{e^2}{c}\vec{A}_p(t)N_e
-\frac{2e}{V_{cell}} \sum_{\vec{k}}\Re\Bigg[D^{\vec{k}*}\vec{P}^{\vec{k}}e^{iS(t)} \nonumber\\
&+&D^{\vec{k}*}C^{\vec{k}}  \left(\frac{1}{m_c}+1\right) \left(\vec{k}+\frac{e}{c}\vec{A}(t)\right)\Bigg],
\end{eqnarray}

\subsubsection{Parallel configuration}
Next, we describe the modulation of the optical properties by the electron excitations.
In this description, we assume the response of diamond to be a typical insulator.
The pump and probe lasers are defined as vector potential fields
\begin{eqnarray}
\vec{A}(t)&=&\vec{A}_0e^{-t^2/\tau_{pump}^2}\cos\Omega t,\\
\vec{A}_p(t)&=&\vec{A}_{p0}e^{-(t-T_p)^2/\tau_{probe}^2}\sin\omega_p (t-T_p).
\end{eqnarray}
The pump pulse has a duration $\tau_{pump}$ set to be $13.5$~fs, whereas the probe pulse has a duration $\tau_{probe}$ set to be 250~as; 
as their polarizations are parallel.
The probe frequency $\omega_p$ is set to be  the optical band gap of the diamond ($B_g=7$~eV), and $\Omega$ is set to be 1.55~eV.
In general, the attosecond pulses used in the experiments are an energy region of 30 to several 100 eV \cite{Chen12,Lucchini16}.
The frequency assumed in this study is much lower than these attosecond pulses.
This assumption is focused on the conceptual understanding, and  
it may be possible to access the experiment by replacing it with a response of a higher lying bands \cite{Lucchini16}.
The effective mass of the conduction band 0.5$m_e$ is used as the reduced mass ($\mu$).  
We assume that $\vec{P}^{\vec{k}}$ does not depend on $\vec{k}$, and that $|\vec{P}^{\vec{k}}|^2$ can be approximated by the Kane's model~\cite{Kane}, 
\begin{equation}
\label{EQ:Kane}
|\vec{P}^{\vec{k}}|^2=m_e^2B_g/4\mu.
\end{equation}
Because the probe light has a peak at time $T_p$, $\sigma(\omega)$ changes to $\sigma(\omega,T_p)$.

We describe the $k$-space using cylindrical coordinates, ($k_r$, $\phi$, $k_z$) and define the space to a cylinder
$0\le k_r \le 0.54$ and $-1.48\le k_z \le 1.48$ given in atomic units (a. u.).
The $k_z$ is parallel to the $\vec{A}_0$.
We prepare a sufficiently large $k_z$ to describe the oscillating wave function under the pump laser field.
The electronic current $J(t)$ is calculated in finite cylinder  $0<k_r \le 0.54$ and $-0.54\le k_z+eA(t)/c \le 0.54$ a.u..
We discretize the $k_r$ and $k_z$ using the mesh size $dk=0.044$ a. u..
The evolution of time proceeds in time steps of $dt=0.08$ a. u., which must be sufficiently small to describe the oscillation of the energy phase of the wave functions.

\begin{figure} 
\includegraphics[width=90mm]{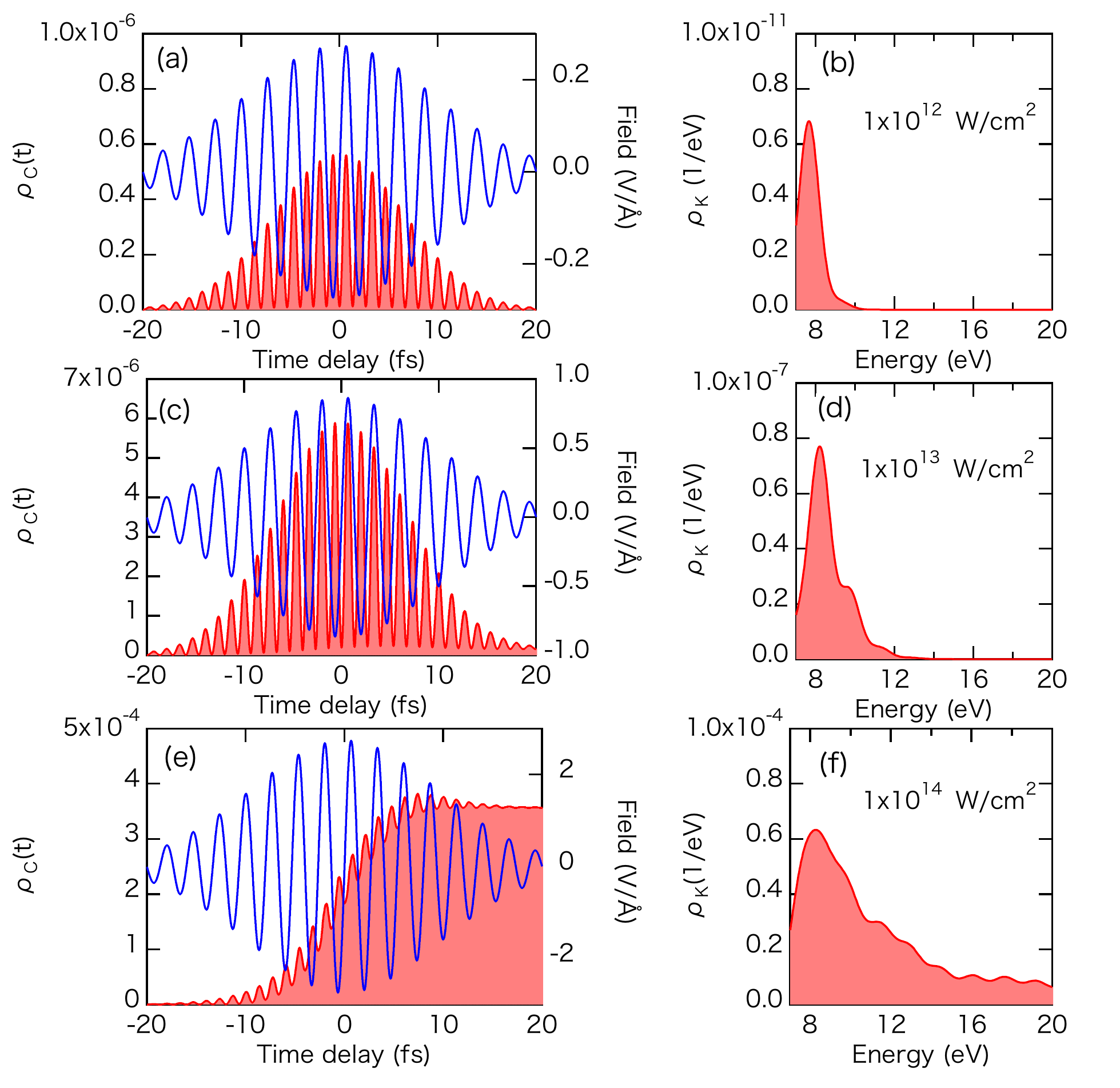} 
\caption{\label{fig1} 
Time-evolution of $\rho_C(t)$ (red line), and the electric field (blue line).
 The pump laser has an intensity of  (a) $1\times10^{12}$~W/cm$^2$, (c) $1\times10^{13}$~W/cm$^2$ and (e) $1\times10^{14}$~W/cm$^2$.
 ((b), (d) and (f)) Energy-gap dependence of the excited electron density after pump excitation at $k=|\vec{k}|$.}
\end{figure}

\begin{figure} 
\includegraphics[width=90mm]{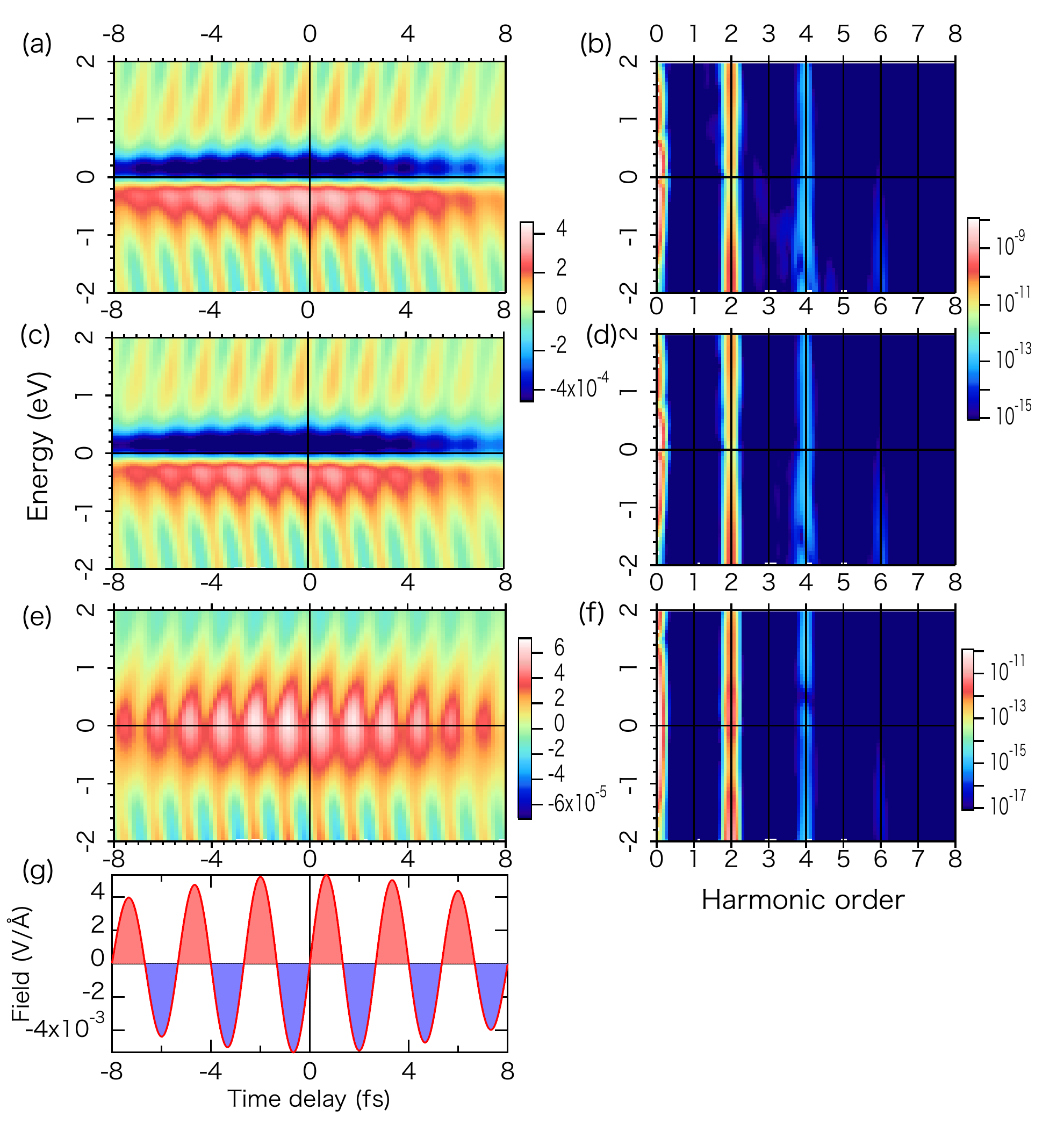} 
\caption{\label{fig2} SCM under a pump intensity of $1\times10^{12}$~W/cm$^2$.
The polarization of the pump and probe lights is parallel.
  (a) Full calculation of the time-evolution of  $ \Re \delta \sigma(\omega,T_p)$. 
  The ordinate represents the energy from $B_g$. 
(b) Fourier components of (a) in a logarithmic scale.  
(c) $\Re\sigma_{DFKE}(\omega,T_p)$.
(d)  Fourier components of (c) in a logarithmic scale.
(e) and (f) show $\Re\sigma_{ex}(\omega,T_p)$ and its Fourier components, respectively.
(g), Applied electric field.} 
 \end{figure}
 \begin{figure} 
\includegraphics[width=90mm]{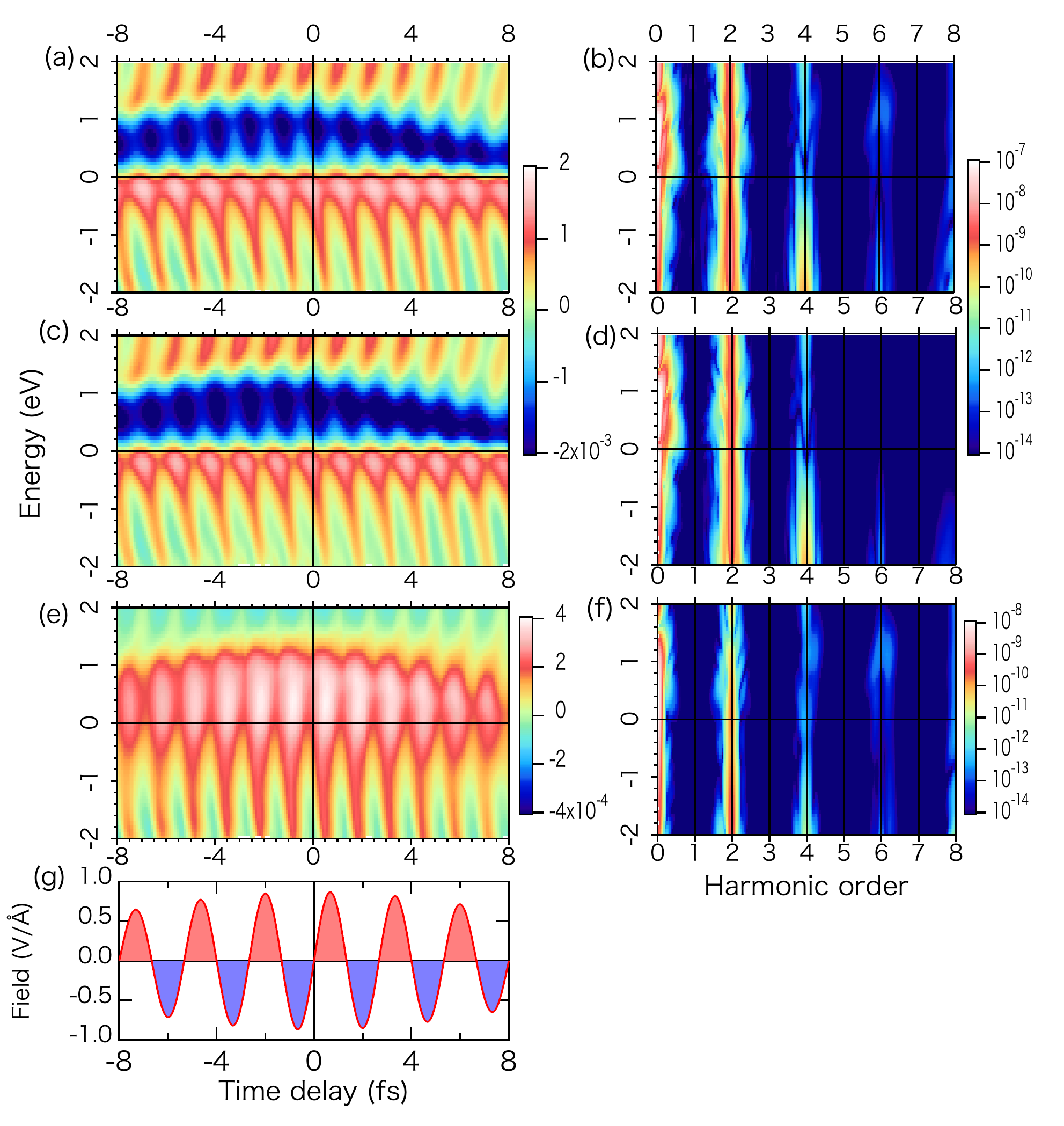} 
\caption{\label{fig3}  SCM under a pump intensity of $1\times10^{13}$~W/cm$^2$.
The description of the each panel corresponds to that in Fig.~\ref{fig2}.} 
 \end{figure}
 
\begin{figure} 
\includegraphics[width=90mm]{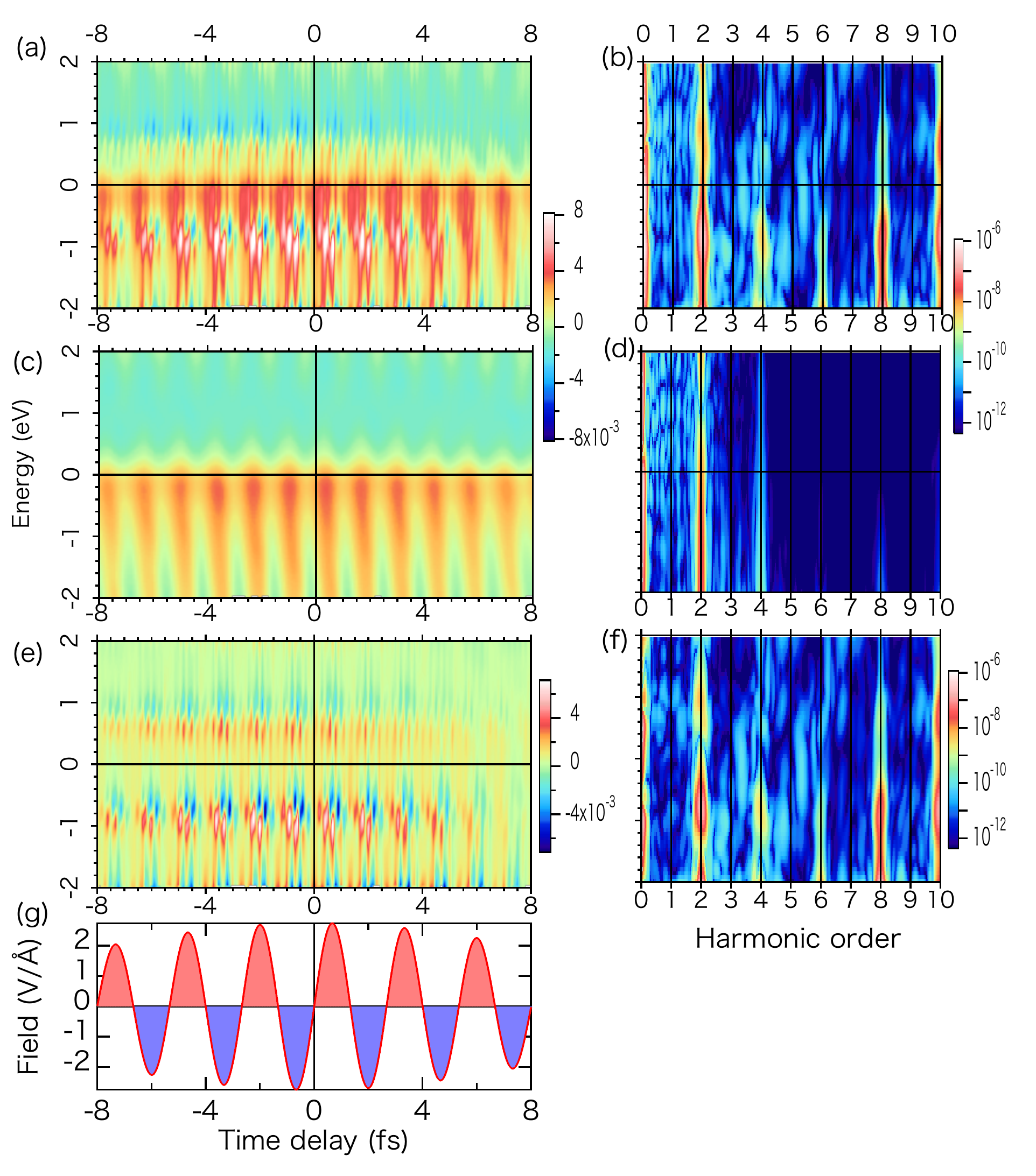} 
\caption{\label{fig4}  SCM under a pump intensity of $1\times10^{14}$~W/cm$^2$.
The description of the each panel corresponds to that in Fig.~\ref{fig2}.} 
 \end{figure} 
 \begin{figure} 
\includegraphics[width=90mm]{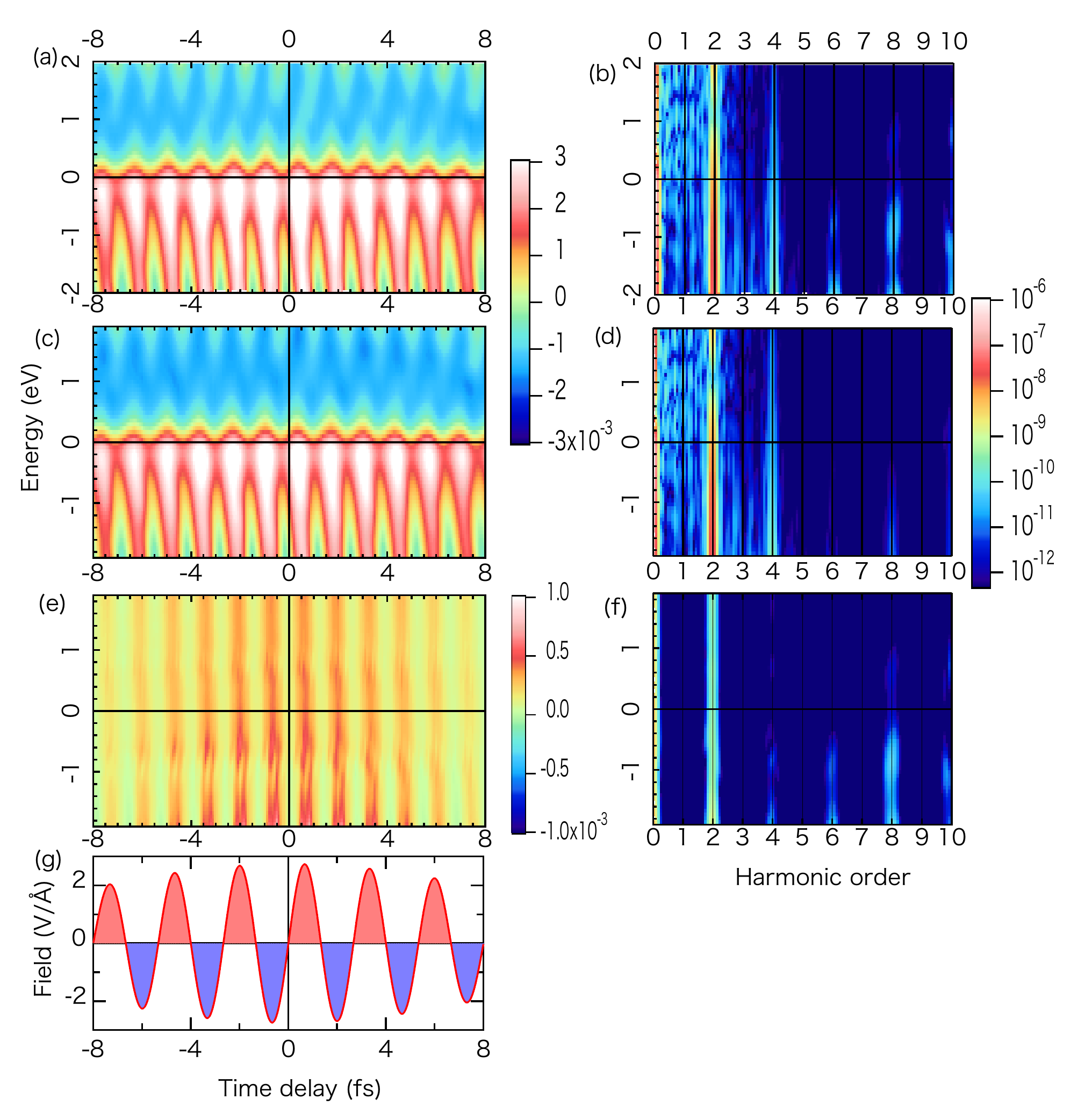} 
\caption{\label{fig5}  SCM under a pump intensity of $1\times10^{14}$~W/cm$^2$.
The polarization of the pump and probe beam is orthogonal, and the description of each panel corresponds to that in Fig.~\ref{fig2}.}
 \end{figure} 
 
Figure~\ref{fig1} desplays the electric field and the electron occupation in the conduction band, $\rho_C(t)=\sum_{\vec{k}}|C^{\vec{k}}(t)|^2/V_{cell}$, 
as a function of time.
The data in Fig.~\ref{fig1}(a) corresponds to a pump beam intensity of $1\times 10^{12}$~W/cm$^2$. 
Occupancy $\rho_C(t)$ exhibits an oscillation with the same frequency as that of the electric field, and indicates negligible electron excitation following the laser pulse. 
Figure~\ref{fig1}(c) and (e) presents results for $1\times 10^{13}$~W/cm$^2$ and $1\times 10^{14}$~W/cm$^2$, respectively. 
Figure~\ref{fig1}(c) shows small electron excitations at 20~fs. 
Because significant electron excitation occurs, a stepwise increase in $\rho_C(t)$ is seen for the intensity $1\times 10^{14}$~W/cm$^2$\cite{Schultze14}. 


The density of the excited electron at $|\vec{k}|$ ($\rho_k$) post excitation is shown in Fig.~\ref{fig1} (b), (d) and (f).
The abscissa of Fig.~\ref{fig1} (b), (d) and (f) gives the energy gap at $|\vec{k}|$.
For $1\times 10^{12}$~W/cm$^2$,  the electron excitation occurs at the band edge.
In contrast, excitations over a broad range in energy occur with a pump intensity of $1\times 10^{14}$~W/cm$^2$.
The small shoulders and oscillation in $\rho_k$ [Fig.~\ref{fig1} (d) and (f)] indicate the multi-photon excitations.

Figure~\ref{fig2} (a) shows the modulation in conductivity, which
corresponds to the difference in the real-part of  $\sigma(\omega,T_p)$ from the conductivity without the pump laser
 $\sigma_0(\omega)$ with all component in Eq.~(\ref{eq:Cur}). 
The Fourier components of the time-dependent modulation, $|F[\Re\delta\sigma(\omega,T_p)]|^2$, is shown in Fig.~\ref{fig2} (b).
The ordinate corresponds to the energy from $B_g$, and the abscissa corresponds to the time delay $T_p$.
The oscillation in $\Re\delta\sigma(\omega,T_p)$ is similar to the typical behavior observed with Tr-DFKE~\cite{otobe16,otobe16-2,otobe16-3,Lucchini16,Uchida16}, which features oscillations with frequencies corresponds to the even harmonics of the $\Omega$ stemming from the symmetry of the system \cite{otobe16,Uchida16}.
 
Figure~\ref{fig2}(c) and (d) presenst the $\Re\sigma_{DFKE}(\omega,T_p)$ and its Fourier transformation $|F[\Re\sigma_{DFKE}(\omega,T_p)]|^2$, respectively. 
The SCM ignoring electron excitation [Fig.~\ref{fig2}(c) and (d)] and the full calculation [Fig.~\ref{fig2}(a) and (b)] show almost identical results and  indeed are indistinguishable.
 Figure~\ref{fig2} (e) shows the $\Re\sigma_{ex}(\omega,T_p)$, which is two orders of magnitude smaller than $\Re\sigma_{DFKE}(\omega,T_p)$.
 Although the electron excitation with time  breaks system periodicity, which was assumed to be an essential requirement for DFKE, 
 the Tr-DFKE is still the dominant effect at this pump intensity.
  However, Fig.~\ref{fig2} (e)  indicates an enhancement in Tr-DFKE (Fig.~\ref{fig2} (c)).
  This enhancement can also be seen in the Fourier transformation of $\Re\sigma_{ex}(\omega,T_p)$; see Fig.~\ref{fig2} (e).
 
 From the results for $1\times 10^{13}$~W/cm$^2$ (Figure~\ref{fig3}), the enhancement of the Tr-DFKE signal by the electron excitation becomes more significant.
 Fig.~\ref{fig3} (b) presents the relatively intense 6th and 8th harmonics,  which are negligible in  Fig.~\ref{fig3} (d).
 The $\Re\sigma_{ex}(\omega,T_p)$ (Fig.~\ref{fig3} (e)) and  its Fourier transformation [Fig.~\ref{fig3} (f)] indicates that 
 the photoabsorption and the even-harmonics are enhanced by the electron excitation. 
 
 At higher pump intensities, we expect that more of the nonlinear components of  $C^{\vec{k}}$ yield higher-order contribution 
 to $|F[ \Re\delta\sigma(\omega,T_p)]|^2$.
 Figure~\ref{fig4} shows results for a pump intensity of $1\times10^{14}$~W/cm$^2$.
While the $\Re\sigma_{DFKE}(\omega,T_p)$ shows the smooth oscillation with respect to the pump field (Fig.~\ref{fig4}(g)), 
 the $\Re\delta\sigma(\omega,T_p)$ shows intense and significantly fast oscillation.

The effect of electron excitation can be seen in the Fourier transformation.
The Fourier components of the spectra of Fig.~\ref{fig4} (a) and (c) are shown in Figs.~\ref{fig4} (b) and (d), respectively.
Whereas $|F[\Re\delta\sigma_{DFKE}(\omega,T_p)]|^2$ without electron excitations has weak harmonics above 4th harmonics, 
a full calculation  shows intense harmonics above 3-rd harmonics.
Figure~\ref{fig4} (f) shows the Fourier transformation of Fig.~\ref{fig4} (e), which shows energy dependent odd- and even-harmonics.
The 8-th and 10-th harmonics are enhanced significantly.
 Therefore, for $1\times10^{14}$~W/cm$^2$, electron excitations enhance the harmonics of $\sigma_{DFKE}$ and generates odd harmonics. 

Whereas beam intensity  $1\times10^{12}$~W/cm$^2$ and $1\times10^{13}$~W/cm$^2$ 
show  enhancements of even harmonics by the electron excitation,
intensity $1\times10^{14}$~W/cm$^2$ exhibits a odd-harmonics.
The even-harmonics modulation has been reported for symmetric materials \cite{Lucchini16,Uchida16}.
In these experiments, the pump laser is set to prevent the electron excitations.
This phenomenon can be attributed to the interaction of electrons excited by the pump and probe pulse.
We shall address the origin of the even- and odd-harmonic modulation in section \ref{harmonics}.

\subsubsection{Orthogonal configuration}
The polarization dependence of SCM is notable when the polarization of the probe light is orthogonal to the pump laser (Figure~\ref{fig4}).
 For the orthogonal configuration, the terms $\vec{A}_p\cdot\vec{A}$ in Eq.~(\ref{eq:Coe_D_para}) and $D^{\vec{k}*}C^{\vec{k}}  \left(\frac{1}{m_c}+1\right) \frac{e}{c}\vec{A}$ in Eq.~(\ref{eq:Cur_para}) are ignored.
 Because the system does not have cylindrical symmetry, we descritize the angle $\phi$ into 64.
 
 Whereas for the parallel configuration exhibits large modulation by the electron excitation,
 $\Re \sigma_{ex}(\omega,T_p)$ [Fig.~\ref{fig5} (e)] is one-third magnitude smaller than  $\Re \sigma(\omega,T_p)$ [Fig.~\ref{fig5} (a)].
 The  Fourier component of $\delta\Re\sigma(\omega,T_p)$ [Fig.~\ref{fig5} (a)] shows the enhancement of the even-harmonics compared with that of $\Re \sigma_{DFKE}(\omega,T_p)$.
 The odd-harmonic modulation for parallel configuration disappears, and the enhancement of even-harmonics decreases for orthogonal configuration.

 \subsubsection{Origin of the harmonics}
 \label{harmonics}
 The origin of the harmonic-order in $|F[\Re\delta\sigma(\omega,T_p)]|^2$ can be understood  
 by changing the pump laser mode from a pulsed to a continuous wave, $\vec{A}(t)=\vec{A}_0\cos\Omega t$.
 The relative phase $e^{iS(t)}$ can then be expanded using the generalized Bessel function \cite{Reiss03} as
 \begin{equation}
 e^{iS(t)}=\sum_l\exp\Bigg[i \Bigg(B_g+U_p+\frac{k^2}{2\mu}+l\Omega\Bigg)t \Bigg]J_l(\alpha,\beta),
 \end{equation}
 where $U_p$ is the ponderomotive energy,  $J_l(\alpha,\beta)$ is the $l$-th order generalized Bessel function 
 with $\alpha=ekA_0\cos\theta/\mu c \Omega$,  and $\beta=e^2A_0^2/8\mu c^2\Omega $; here $\theta$ is the angle between $\vec{k}$ and $\vec{A}_0$ \cite{otobe16}.
We have then
 \begin{eqnarray}
&&C^{\vec{k}}(t)=-e\int^t dt'\frac{\vec{E}(t')\cdot\vec{P}^{\vec{k}} }{B_g+\frac{\left(\vec{k}+\frac{e}{c}\vec{A}(t')\right)^2}{2\mu}}e^{iS(t)}\nonumber\\
&=&\frac{e}{c}\frac{\vec{A}_0\cos\Omega t\cdot\vec{P}^{\vec{k}}}{B_g+\frac{\left(\vec{k}+\frac{e}{c}\vec{A}\cos\Omega t\right)^2}{2\mu}}e^{iS(t')}\nonumber\\
&-&i\frac{e}{c}\vec{A}_0\cdot\vec{P}^{\vec{k}}\int_{-\infty}^{\infty} ds\Theta(s)\cos\Omega (t-s) \nonumber\\
&\times&\sum_l\exp\Big[i \kappa_l (t-s) \Big]J_l(\alpha,\beta),
\end{eqnarray}
where $s=t-t'$, $\kappa_l=B_g+U_p+\frac{k^2}{2\mu}+l\Omega$, and $\Theta(s)$ is the Heaviside function introduced to satisfy the causality.

If the $B_g$ is large, the  $C^{\vec{k}}$ can be approximated by 
 \begin{eqnarray}
 \label{C_CW}
&&C^{\vec{k}}(t)
\approx-\frac{i\pi e}{ c}\vec{A}_0\cdot\vec{P}^{\vec{k}}\sum_{l} e^{i\kappa_{l}t} \nonumber\\
&\times& \left(\pi \delta(\Omega-\xi_k/l)+\frac{i}{\Omega-\xi_k/l}\right) \nonumber\\
&\times&(J_{l+1}(\alpha,\beta)+J_{l-1}(\alpha,\beta)),
\end{eqnarray}
where $\xi_k=B_g+U_p+\frac{k^2}{2\mu}$.
 The index $l$ gives the harmonic order of the frequency $l\Omega$ for the oscillation. 
 The coefficient $D^{\vec{k}}$ and the current $\vec{J}(t)$ can also be written as:
 \begin{eqnarray}
&&D^{\vec{k}}(t)
=-\frac{ie}{ c}\int dt'\vec{P}^{\vec{k}}\cdot\vec{A}_p(t')
\sum_l e^{i\kappa_l t'}J_l(\alpha,\beta)\nonumber\\
&-&\frac{ie}{ c} \left(\frac{1}{ m_c}+1 \right)\int dt' C^{\vec{k}}(t')\nonumber\\
&\times&\vec{A}_p(t') \cdot\left( \vec{k}+\frac{e}{ c}\vec{A}_0\cos\Omega t'\right),
\end{eqnarray}
and
\begin{eqnarray}
\label{cur_CW}
\vec{J}(t)&=&\vec{J}_P(t) -\frac{e^2}{c}\vec{A}_p(t)N_e-\frac{2e}{V_{cell}} \sum_{\vec{k}}\Re\Bigg[D^{\vec{k}*}\vec{P}^{\vec{k}}\nonumber\\
&\times&\sum_l e^{i\kappa_l  t}J_l(\alpha,\beta)\nonumber\\
&+&D^{\vec{k}*}C^{\vec{k}}  \left(\frac{1}{m_c}+1\right) \left(\vec{k}+\frac{e}{c}\vec{A}_0\cos\Omega t\right)
  \Bigg].
\end{eqnarray}

If the probe pulse is extremely short and can be approximated as $\vec{A}_p(t)=\vec{A}_{p0}\delta(t-T_p)$, $D^{\vec{k}}(t)$ becomes simply
 \begin{eqnarray}
&&D^{\vec{k}}(t=T_p)
=-\frac{ie}{ c}\vec{P}^{\vec{k}}\cdot\vec{A}_{p0}
\sum_l e^{i\kappa_l T_p}J_l(\alpha,\beta)\nonumber\\
&-&\frac{\pi e^2}{ c^2} \left(\frac{1}{ m_c}+1 \right)\vec{A}_{p0} \cdot\left( \vec{k}+\frac{e}{ c}\vec{A}_0\cos\Omega T_p\right)\nonumber\\
&\times&\vec{A}_0\cdot\vec{P}^{\vec{k}}\sum_{l} e^{i\kappa_{l}T_p} 
 \left(\pi \delta(\Omega-\xi_k/l)+\frac{i}{\Omega-\xi_k/l}\right) \nonumber\\
&\times&(J_{l+1}(\alpha,\beta)+J_{l-1}(\alpha,\beta))
\end{eqnarray}
Because $\cos \Omega t$ change the generalized Bessel function $J_{l}(\alpha,\beta)$ to $(J_{l+1}(\alpha,\beta)+J_{l-1}(\alpha,\beta))/2$, 
the coefficient $D^{\vec{k}}$ becomes
 \begin{eqnarray}
 \label{D_CW_last}
&&D^{\vec{k}}(t=T_p)
=-\frac{ie}{ c}\vec{P}^{\vec{k}}\cdot\vec{A}_{p0}
\sum_l e^{i\kappa_l T_p}J_l(\alpha,\beta)\nonumber\\
&-&\frac{\pi e^2}{ c^2} \left(\frac{1}{ m_c}+1 \right)(\vec{A}_{p0} \cdot \vec{k})(\vec{A}_0\cdot\vec{P}^{\vec{k}})\nonumber\\
&\times&\sum_{l} e^{i\kappa_{l}T_p} 
 \left(\pi \delta(\Omega-\xi_k/l)+\frac{i}{\Omega-\xi_k/l}\right) \nonumber\\
&\times&(J_{l+1}(\alpha,\beta)+J_{l-1}(\alpha,\beta))\nonumber\\
&-&\frac{\pi e^3}{ 2c^3} \left(\frac{1}{ m_c}+1 \right)(\vec{A}_{p0} \cdot \vec{A}_0)(\vec{A}_0\cdot\vec{P}^{\vec{k}})\nonumber\\
&\times&\sum_{n=-1,1}\sum_{l} e^{i\kappa_{l}T_p} 
 \left(\pi \delta(\Omega-\xi_k/(l+n))+\frac{i}{\Omega-\xi_k/(l+n)}\right) \nonumber\\
&\times&(J_{l+1+n}(\alpha,\beta)+J_{l-1+n}(\alpha,\beta)).
\end{eqnarray}

 $J(t)$ contains the components $J_lJ_{l'}$, $J_lJ_{l'\pm1}$, and $J_lJ_{l'\pm2}$.
For each term, $l+l'$,  $l+l'\pm1$, and $l+l'\pm2$ in each components becomes either even or zero from the symmetry of the system.
Therefore, the terms in in $\Re\sigma_{ex}(\omega,T_p)$ that contains $J_lJ_{l'}$ and $J_lJ_{l'\pm2}$ give even- and zero-order harmonics,  
wheseas those that contain $J_lJ_{l'\pm1}$ gives odd-order harmonics.
Because $D^{\vec{k}}(t=T_p)$ has $J_lJ_{l'}$, $J_lJ_{l'\pm1}$, and $J_lJ_{l'\pm2}$, 
all terms in $\vec{J}(t)$ contain $J_lJ_{l'\pm1}$.
If we ignore $C^{\vec{k}}$, all terms containing  $J_lJ_{l'\pm1}$ disappear, and $\Re\sigma(\omega,T_p)$ exhibits only even-harmonics.

The important effect is the path interference between the excited electrons expressed as the term $D^{\vec{k}*}C^{\vec{k}}$, 
that is, interference between the electron excited by pump and probe pulses.
For Tr-DFKE, the order of the generalized Bessel functions coincides with the frequency of $\sigma(\omega,T_p)$.
In contrast, the order of the generalized Bessel functions in $C^{\vec{k}}$ (Eq.~(\ref{C_CW})) is shifted $\pm1$ from the frequency of $C^{\vec{k}}$ because
 it contains $\vec{A}(t)$, which results in the odd-harmonics in the term $D^{\vec{k}*}C^{\vec{k}}(k_z+eA(t)/c)$. 
 In other words, the odd-harmonics reflects the asymmetric distribution  of $C^{\vec{k}}$ with respect to the oscillating $\Gamma$-point under the pump field, $eA(t)/c$.

For an orthogonal configuration, the contribution of the  third term in Eq.~(\ref{D_CW_last}) vanishes.
After the integration in  the $\vec{k}$-space, the contribution of the first term of Eq.~(\ref{D_CW_last}) vanishes in  $D^{\vec{k}*}C^{\vec{k}}$.
In contrast, the second term of Eq.~(\ref{D_CW_last}) vanishes in the term $D^{\vec{k}*}\vec{P}^{\vec{k}}$ in the Eq.~(\ref{cur_CW})
Therefore, the term $D^{\vec{k}*}\vec{P}^{\vec{k}}$ corresponds to Tr-DFKE.
Because the direction of the motion of the wave packets produced by the pump and probe is orthogonal, its effect is relatively small.
This is the reason why the effect of the electron excitation becomes weak in orthogonal configuration.

\subsection{Cosine band}
The parabolic-band model provides the simplest approximation of the band structure.
An alternative option is the cosine band model, which may take into account the non-harmonic structure of the real materials.

We initialize the CB and VB calculation with the parameter settings for diamond, that is,  
\begin{eqnarray}
\varepsilon_c^{k}(t)&=&\tilde B_g+\frac{\Delta E_c}{2}\left(\cos dk+1\right)\\
\varepsilon_v^{k}(t)&=&\frac{\Delta E_v}{2}\left(\cos dk-1\right)\\
P_{cc}^{k}&=&-\frac{\Delta E_c d}{2}\sin d k,
\end{eqnarray}
where
$\tilde B_g=5$~eV is the indirect band gap, $\Delta E_c=2$~eV and  $\Delta E_v=7$~eV are the widths of the conduction and valence bands, and $d=3.567$~\AA~is the lattice constant.
We change the parameters of the pump laser field to  $\Omega=0.5$~eV, $\tau=10$~fs to compare results with the previous works employing time-dependent density functional theory (TDDFT) \cite{otobe16}. 
We assume a one-dimensional band in the following discussion.

\begin{figure} 
\includegraphics[width=90mm]{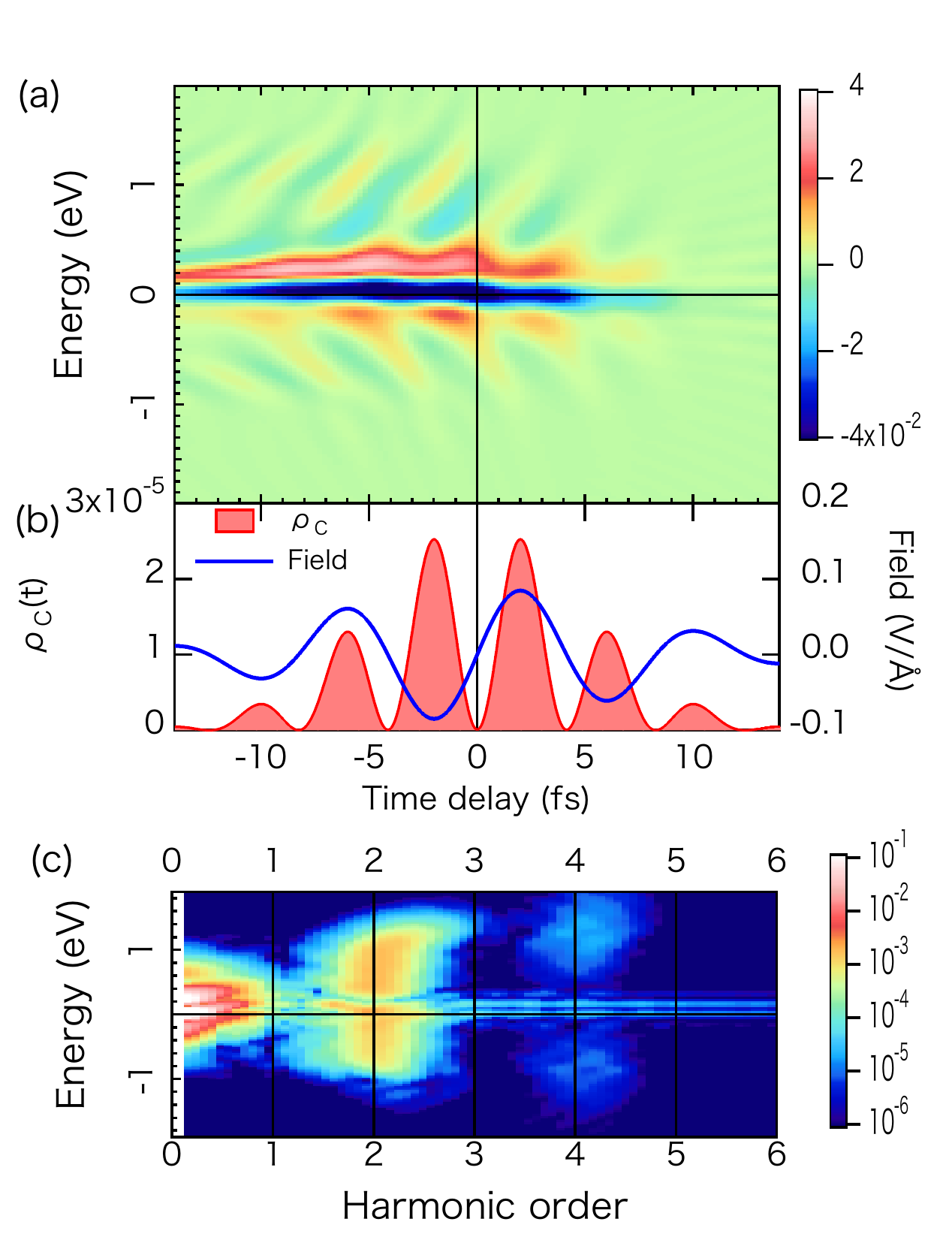} 
\caption{\label{fig6} 
 SCM under a pump intensity of $1\times10^{11}$~W/cm$^2$ analized using the cosine band model;
 (a) SCM with full component; (b) time-evolution of $\rho_c(t)$ (red) and pump field (blue);
 (c) Fourier transformation of (a).}
\end{figure}

Because the calculation for the Tr-DFKE with the cosine band  model has not been reported, we shall confirm that this model reproduces the usual Tr-DFKE signal qualitatively.
Figure~\ref{fig6} shows results for a pump  intensity of $1\times10^{11}$~W/cm$^2$. 
Figure~\ref{fig6} (a) presents the $\delta \Re \sigma(\omega,T_p)$ for the system subject to the pump laser field [Fig.~\ref{fig6}(b); blue line]. 
The electron excitation is negligible with this pump intensity [Fig.~\ref{fig6} (b); red shaded area].  
The frequency-dependent oscillation of $\delta \Re \sigma(\omega,T_p)$, which is the significant feature of the Tr-DFKE \cite{otobe16}. 
A blue shift also appears in band gap because of ponderomotive energy, which results from the decrease in $ \Re \sigma(\omega,T_p)$ above the band gap.

$F[\delta \Re \sigma(\omega,T_p)]$ [Fig.~\ref{fig6} (c)] indicates that the oscillation of $\delta \Re \sigma(\omega,T_p)$ has the even-order harmonics,
which are also a feature of Tr-DFKE \cite{otobe16,Lucchini16,Uchida16}.
From these results, the cosine band model can also describes the Tr-DFKE for weak pump fields in SCM.

\begin{figure} 
\includegraphics[width=90mm]{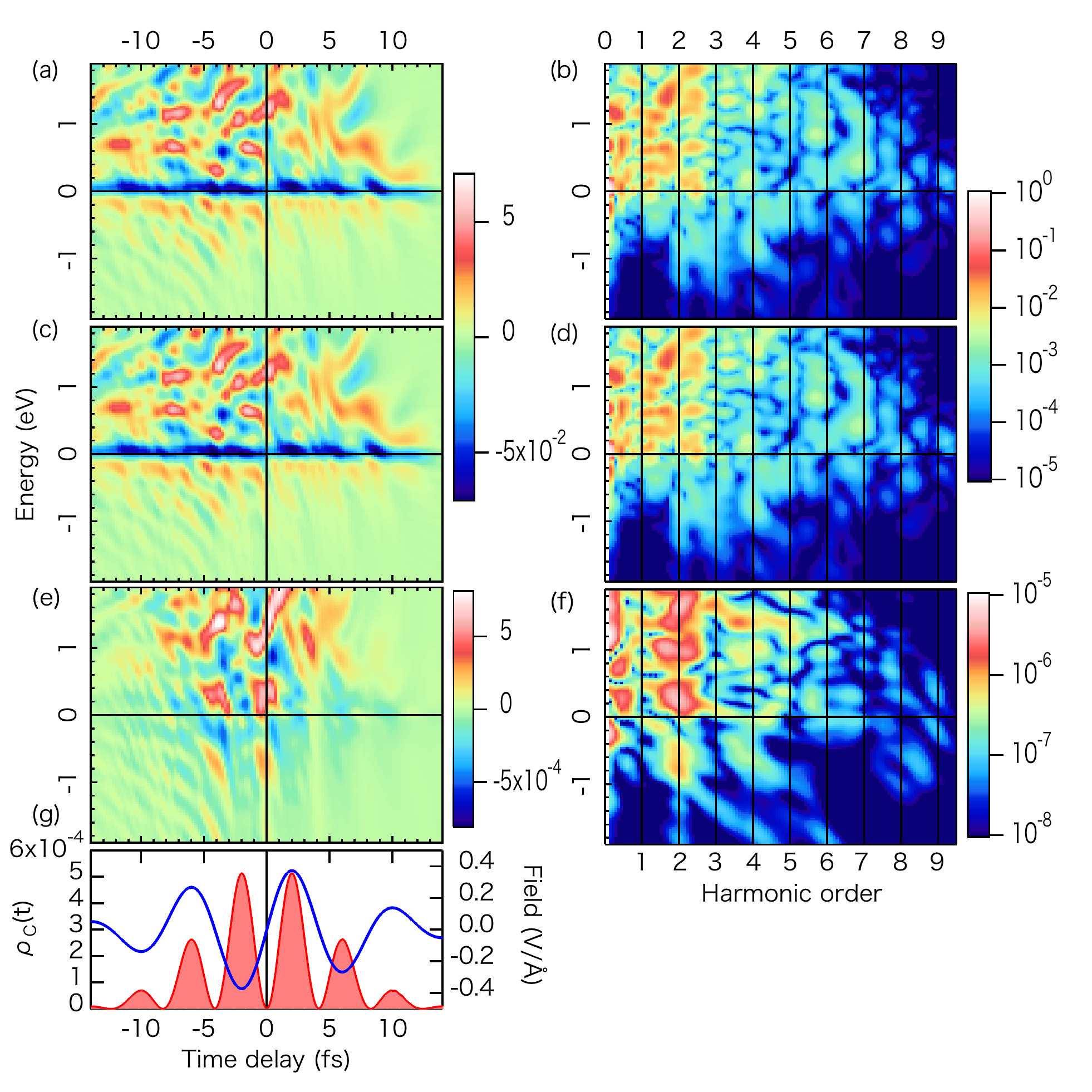} 
\caption{\label{fig7}  SCM under a pump intensity of $2\times10^{12}$~W/cm$^2$ analyzed using cosine band model.
  (a) Full calculation of the time-evolution of  $ \Re \delta \sigma(\omega,T_p)$. 
  The ordinaterepresents the energy from the $B_g$. 
(b) Fourier components of (a) in a logarithmic scale.  
(c) $\Re\sigma_{DFKE}(\omega,T_p)$.
(d)  Fourier components of (c) in a logarithmic scale.
(e) and (f) show $\Re\sigma_{ex}(\omega,T_p)$ and its Fourier components, respectively.
(g), Applied electric field (bleu) and $\rho_C(t)$ (red).} 
 \end{figure}

Figure~\ref{fig7} shows results for laser intensity $2\times10^{12}$ W/cm$^2$.
According to our previous result using the time-dependent density functional theory, 
a clear oscillation in $\Re\sigma_{DFKE}$ with respect to the pump field is seen in this pump intensity regime (Fig. 2 in \cite{otobe16}).
However, the cosine band model does not reproduce the Tr-DFKE feature even when electron excitations are ignored [Fig.~\ref{fig7} (c) and (d)].

\begin{figure} 
\includegraphics[width=90mm]{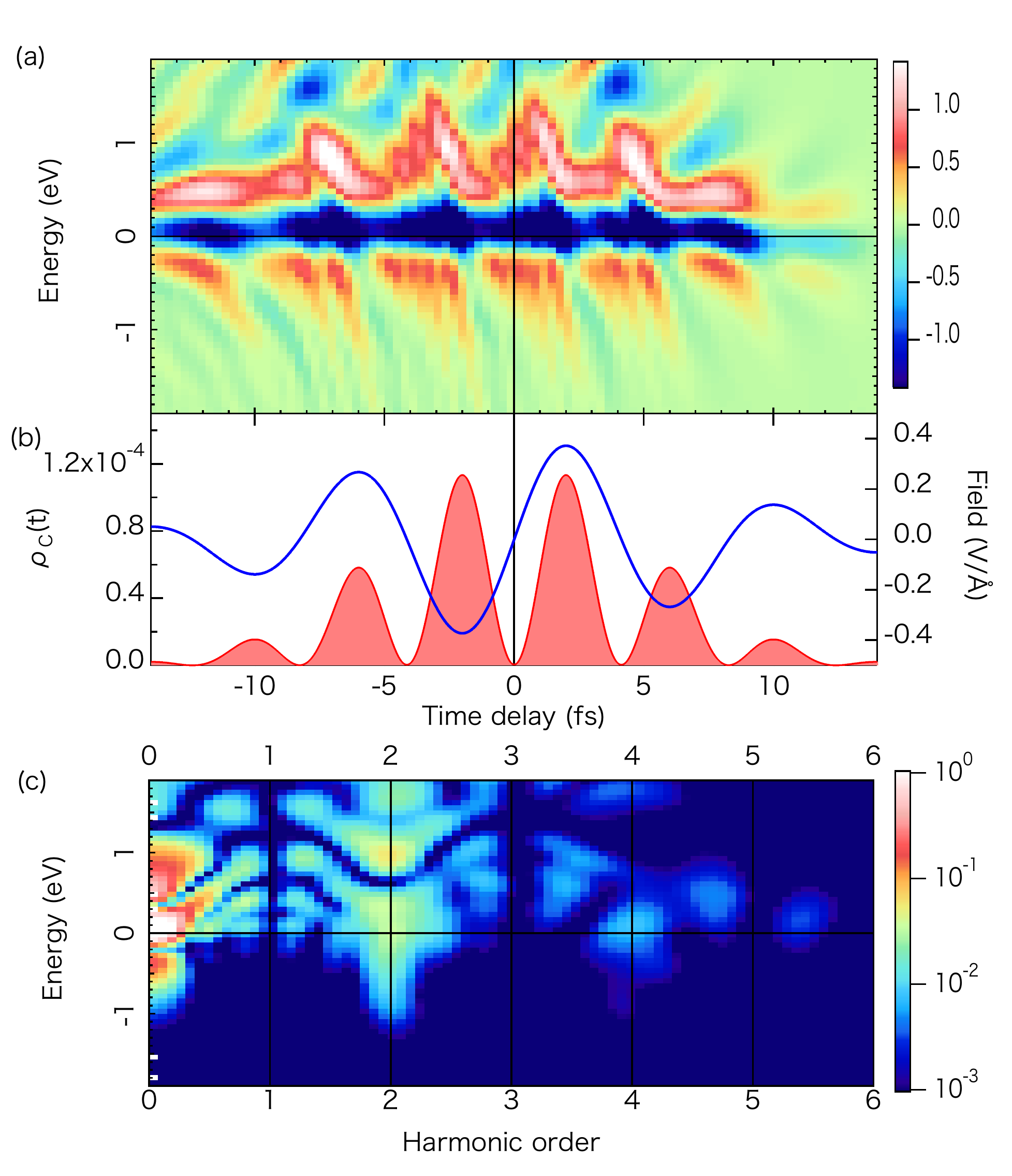} 
\caption{\label{fig8} 
 SCM under a pump intensity of $2\times10^{12}$=W/cm$^2$ analyzed using one-dimensional parabolic band.
 (a) SCM with full component; (b) Time-evolution of $\rho_c(t)$ (red) and pump field (blue);
 (c) Fourier transformation of (a).}
\end{figure}

One possible reason for this breakdown is the effect of one-dimensional assumption.
Assuming a one-dimensional parabolic band, the results for a pump intensity of $2\times10^{12}$~W/cm$^2$ [Fig.~\ref{fig8}(a)] shows oscillation following the pump field.
The even-harmonic spectrum in the $F[\delta \Re \sigma(\omega,T_p)]$ is also reasonably reproduced [Fig.~\ref{fig8} (c)]
This result indicates that the cosine band model overestimates the non-parabolic structure of both the CB and VB.
In real materials, the energy gaps in CB and VB are sufficiently small compared with the photon energy,  ponderomotive energy, and/or $E_0 d$. 
Therefore, the electron wave function can  be accelerated as a quasi-free electron.
In such instances, the inter-band transition in CB and VB assume some importance at extremely intense pump-laser intensities, and the model should be expanded to a multi-band system. 
The importance of the multi-band system is also reported for high-harmonic generation  in dielectrics by Hawkins \textit{et al.} \cite{Hawkins15} and Ikemachi \textit{et al.} \cite{Ikemachi17}.

 
\section{summary}
In summary, we presented an analytical theory
for SCM, which includes electron excitations induced by a pump-laser field.
Although our results indicate that the Tr-DFKE is a fundamental and robust effect in the SCM,
we found that the electron excitation enhances the even-harmonic modulation and generates odd-harmonic modulation in optical properties.
The relative polarization of pump and probe pulse also changes the SCM significantly when the electron excitation occurs.
For orthogonal configuration, the effect is small modulation compared with that for parallel configulation.
 We also found that the parabolic band model is more suitable than cosine the band model, indicating that the electrons
  move as the quasi-free particles under the intense laser fields.

 The modulation of Tr-DFKE using the electron excitations and the polarization is the new aspect of the material control in sub-cycle time-domain.
  Our results indicate a new approach in optimizing the ultrafast sub-cycle switching of material properties, 
which is a key phenomenon in so-called peta-Hertz engineering \cite{Schultze13,Mashiko16}. 
 
\section*{Acknowledgement}
This work was supported by  JSPS KAKENHI (Grants Nos. 15H03674 and 17K05089). 
Numerical calculations were performed on the SGI ICE X supercomputer at 
the Japan Atomic Energy Agency (JAEA).

\appendix
\section{Coefficients}
In this appendix, we would like to present the derivation of the coefficients $C^{\vec{k}}(t)$ and $D^{\vec{k}}(t)$.

The time evolution of the wave function of valence band, $u_{v,\vec{k}}$, is given by
\begin{equation}
\label{TDSE1_app}
i\frac{\partial u_{v,\vec{k}}(t) }{\partial t}=H(t)u_{v,\vec{k}}(t),
\end{equation}
where $H(t)$ is the Hamiltonian including the pump laser field,
\begin{equation}
H(t)=\frac{1}{2}\left(\vec{p}+\vec{k}+\frac{e}{ c}\vec{A}(t)\right)^2+V(\vec{r}).
\end{equation}
We assume that  the  $u_{v,\vec{k}}$ can be expanded by the Houston function and the coefficient $C^{\vec{k}}(t)$,
\begin{equation}
u_{v,\vec{k}}(\vec{r},t)=w_{v,\vec{k}}(\vec{r},t)+C^{\vec{k}}(t)w_{c,\vec{k}}(\vec{r},t).
\end{equation}
Then the Eq.~(\ref{TDSE1_app}) reads as
\begin{eqnarray}
&&i\frac{\partial}{\partial t} u_{v,\vec{k}}(\vec{r},t)= i\frac{\partial}{\partial t} \left\{w_{v,\vec{k}}(\vec{r},t)+C^{\vec{k}}(t)w_{c,\vec{k}}(\vec{r},t)\right\}\nonumber\\
&=&i\left[\frac{\partial w_{v,\vec{k}}(\vec{r},t)}{\partial t}+\frac{\partial C^{\vec{k}}(t)}{\partial t}w_{c,\vec{k}}(\vec{r},t)+C^{\vec{k}}(t)\frac{\partial w_{c,\vec{k}}(\vec{r},t)}{\partial t}\right]\nonumber\\
&=& i\frac{\partial C^{\vec{k}}(t)}{\partial t}w_{c,\vec{k}}(\vec{r},t)\nonumber\\
&+&\Bigg\{-ie\vec{E}(t)\cdot\frac{\partial u_{v,\vec{k}}^G(\vec{r})}{\partial \vec{k}}\Big|_{\vec{k}+\frac{e}{c}\vec{A}(t)}e^{-i\int^t dt' \varepsilon_{v,\vec{k}}^G(t')}\nonumber\\
&+&\varepsilon_{v,\vec{k}}^G(t)u_{v,\vec{k}+\frac{e}{ c}\vec{A}(t)}^G(\vec{r})e^{-i\int^t dt' \varepsilon_{v,\vec{k}}^G (t')}\Bigg\}\nonumber\\
&+&C^{\vec{k}}(t)\varepsilon_{c,\vec{k}}^G(t)u_{c,\vec{k}+\frac{e}{ c}\vec{A}(t)}^G(\vec{r})e^{-i\int^t dt' \varepsilon_{c,\vec{k}}^G(t') }.
\end{eqnarray}
The equation about the $C^{\vec{k}}(t)$ is obtain by applying $\left< w_{c,\vec{k}}(\vec{r},t) \right|$ from the left,
\begin{eqnarray}
&i&\left< w_{c,\vec{k}}(\vec{r},t) \Bigg|\frac{\partial}{\partial t} u_{v,\vec{k}}(\vec{r},t)\right>=i\frac{\partial C^{\vec{k}}(t)}{\partial t}\nonumber\\
&-&i\left< u_{c,\vec{k}}^G \Bigg| \frac{\partial u_{v,\vec{k}}^G}{\partial \vec{k}}\right> \Big|_{\vec{k}+\frac{e}{c}\vec{A}(t)}e\vec{E}(t)e^{i S(t)} \nonumber\\
&+&C^{\vec{k}}(t)\varepsilon_{c,\vec{k}}^G(t).
\end{eqnarray}
Then the  $C^{\vec{k}}$ is given as
\begin{eqnarray}
\label{eq:C_app1}
&&C^{\vec{k}}(t)\nonumber\\
&=&e\int^t dt' \vec{E}(t')\cdot\left< u_{c,\vec{k}}^G \Bigg| \frac{\partial u_{v,\vec{k}}^G}{\partial \vec{k}}\right>\Big|_{\vec{k}+\frac{e}{c}\vec{A}(t')} e^{i S(t')}~~~~~~\\
\label{eq:C_app2}
&=&-ie\int^t dt' \vec{E}(t')\cdot\vec{d}^{\vec{k}+\frac{e}{c}\vec{A}(t')}e^{i S(t')} \\
\label{eq:C_app3}
&=&-e\int^t dt' \vec{E}(t')\cdot\frac{\vec{P}^{\vec{k}+\frac{e}{c}\vec{A}(t')}}{\varepsilon^G_{c,\vec{k}}(t')-\varepsilon^G_{v,\vec{k}}(t') }e^{i S(t')},
\end{eqnarray}
where
\begin{equation}
\vec{d}^{\vec{k}}=\left< u_{c,\vec{k}}^G \Big| \vec{r}\Big| u_{v,\vec{k}}^G\right>,
\end{equation}
and
\begin{equation}
\vec{P}^{\vec{k}}=\left< u_{c,\vec{k}}^G \Big| \vec{p}\Big| u_{v,\vec{k}}^G\right>.
\end{equation}
The time-evolution of $C^{\vec{k}}(t)$  followd by applying the band structure $\varepsilon^G_{c,\vec{k}}(t)-\varepsilon^G_{v,\vec{k}}(t)$,  
the transition dipole moment, and the transition momentum.
Eqs.~(\ref{eq:C_app1})-(\ref{eq:C_app3}) corresponds to the generalization of the Keldysh theory \cite{Keldysh65,Gruzdev07,McDonald17,Ex}.

The time-dependent Schr\"odinger equation describing the system subject to  pump and probe pulses becomes,
\begin{equation}
i\frac{\partial \tilde{u}_{v,\vec{k}}(\vec{r},t) }{\partial t}=\left[H(t)+\delta H(t)\right]\tilde{u}_{v,\vec{k}}(\vec{r},t),
\end{equation}
where
\begin{equation}
\delta H(t)\approx\frac{e}{c}\left(\vec{p}+\vec{k}+\frac{e}{c}\vec{A}(t)\right)\cdot\vec{A}_p(t).
\end{equation}
The time-dependent wave function $\tilde{u}_{v,\vec{k}}(\vec{r},t)$ is also expanded using $D^{\vec{k}}(t)$,
\begin{equation}
\tilde{u}_{v,\vec{k}}(\vec{r},t)=u_{v,\vec{k}}(\vec{r},t)+ D^{\vec{k}}(t) w_{c,\vec{k}}(\vec{r},t).
\end{equation}

The time-evolution of $D^{\vec{k}}(t)$ is 
\begin{eqnarray}
&&i\frac{\partial \tilde{u}_{v,\vec{k}}(\vec{r},t)}{\partial t}=i \Bigg\{\frac{\partial u_{v,\vec{k}}(\vec{r},t)}{\partial t} \nonumber\\
&+&\left(\frac{\partial D^{\vec{k}}(t)}{\partial t}w_{c,\vec{k}}(\vec{r},t)+D^{\vec{k}}(t)\frac{\partial w_{c,\vec{k}}(\vec{r},t)}{\partial t}\right)\Bigg\} \nonumber\\
&=&(H(t)+\delta H(t))\left(u_{v,\vec{k}}(\vec{r},t)+ D^{\vec{k}}(t) w_{v,\vec{k}}(\vec{r},t)\right)\nonumber\\
&\approx& H(t)u_{n,\vec{k}}(\vec{r},t)+\delta H(t)u_{n,\vec{k}}(\vec{r},t)\nonumber\\
&+&H(t)\sum_v D^{\vec{k}}(t) w_{v,\vec{k}}(\vec{r},t)\\
&=& \frac{\partial u_{n,\vec{k}}(\vec{r},t)}{\partial t}+\delta H(t)u_{n,\vec{k}}(\vec{r},t) \nonumber\\
&+& D^{\vec{k}}(t)\varepsilon_{c,\vec{k}+\frac{e}{ c}\vec{A}(t)} u_{v,\vec{k}+\frac{e}{ c}\vec{A}(\vec{r},t)}^G(\vec{r})e^{iS(t)}.
\end{eqnarray}
In this step, we assume that the probe pulse is weak and the linear term is dominant.

Similar to $C^{\vec{k}}(t)$,  $D^{\vec{k}}(t)$ is given as
\begin{eqnarray}
\label{eq:D_app}
D^{\vec{k}}(t)&=&-\frac{ie}{ c}\int dt' \vec{P}^{\vec{k}+\frac{e}{c}\vec{A}(t')}\cdot\vec{A}_p(t')
e^{ i S(t')}\nonumber\\
&-&\frac{ie}{ c}\int dt' \Big[C^{\vec{k}}(t')\vec{P}^{\vec{k}+\frac{e}{c}\vec{A}(t')}_{cc}\cdot\vec{A}_p(t') \nonumber\\
&+&C^{\vec{k}}(t')\vec{A}_p(t')\cdot\left(\vec{k}+\frac{e}{ c}\vec{A}(t')\right)\Big\}\Big].
\end{eqnarray}
Therefore, $\tilde{u}_{v,\vec{k}}(\vec{r},t)$  and the physical quantities can be calculated using Eqs.~(\ref{eq:C_app3}) and (\ref{eq:D_app})

\section{Current}
The total current $\vec{J}(t)$ is given by the momentum operator $\vec{p}+\vec{k}+\frac{e}{c}\left(\vec{A}(t)+\vec{A}_p(t)\right)$ and wave function 
$\tilde{u}_{v,\vec{k}}(\vec{r},t)=w_{v,\vec{k}}(\vec{r},t)+ C^{\vec{k}}(t) w_{c,\vec{k}}(\vec{r},t)+ D^{\vec{k}}(t) w_{c,\vec{k}}(\vec{r},t)$ as,
\begin{eqnarray}
\label{eq:Cur_app}
&&\vec{J}(t)\nonumber\\
&=&-\frac{e}{V_{cell}}\sum_{\vec{k}}\Re\left<\tilde{u}_{v,\vec{k}}\left|\vec{p}+\vec{k}+\frac{e}{c}\left(\vec{A}(t)+\vec{A}_p(t)\right)\right|\tilde{u}_{v,\vec{k}}\right>\nonumber\\
&=&-\frac{e}{V_{cell}}\sum_{\vec{k}} \Re\Bigg[\left<w_{v,\vec{k}}\left|\vec{p}+\vec{k}+\frac{e}{c}\left(\vec{A}(t)+\vec{A}_p(t)\right)\right|w_{v,\vec{k}}\right>\nonumber\\
&+&2 C^{*\vec{k}}(t)\vec{P}^{\vec{k}+\frac{e}{c}\vec{A}(t)}e^{iS(t)}+2 D^{*\vec{k}}(t)\vec{P}^{\vec{k}+\frac{e}{c}\vec{A}(t)}e^{iS(t)}\nonumber\\
&+&2 D^{*\vec{k}}C^{\vec{k}}(t)\left<w_{c,\vec{k}}\left|\vec{p}+\vec{k}+\frac{e}{c}\left(\vec{A}(t)+\vec{A}_p(t)\right)\right|w_{c,\vec{k}}\right>\nonumber\\
&+& |C^{\vec{k}}(t)|^2\left<w_{c,\vec{k}}\left|\vec{p}+\vec{k}+\frac{e}{c}\left(\vec{A}(t)+\vec{A}_p(t)\right)\right|w_{c,\vec{k}}\right>\nonumber\\
&+& |D^{\vec{k}}(t)|^2\left<w_{c,\vec{k}}\left|\vec{p}+\vec{k}+\frac{e}{c}\left(\vec{A}(t)+\vec{A}_p(t)\right)\right|w_{c,\vec{k}}\right>\Bigg].~~~~~~
\end{eqnarray}
Assuming a weak probe pulse, the terms containing $D^{*\vec{k}}\vec{A}_p(t)$ or $|D^{\vec{k}}(t)|^2$ can be neglected. 
The contribution of the pump pulse ($\vec{J}_p(t)$) to the current is 
\begin{eqnarray}
&&\vec{J}_P(t)=-\frac{e}{V_{cell}}\sum_{\vec{k}} \nonumber\\
&&\Re\left<w_{v,\vec{k}}\left|\vec{p}+\vec{k}+\frac{e}{c}\vec{A}(t)\right|w_{v,\vec{k}}\right>\nonumber\\
&-&\frac{e}{V_{cell}}\sum_{\vec{k}} \Re\Bigg[2 C^{*\vec{k}}(t)\vec{P}^{\vec{k}+\frac{e}{c}\vec{A}(t)}e^{iS(t)}\nonumber\\
&+&|C^{\vec{k}}(t)|^2\left<w_{c,\vec{k}}\left|\vec{p}+\vec{k}+\frac{e}{c}\vec{A}(t)\right|w_{c,\vec{k}}\right>\Bigg].
\end{eqnarray}
Then the $\vec{J}(t)$ can be written as,
\begin{eqnarray}
\label{eq:Cur_app2}
&&\vec{J}(t)
\approx\vec{J}_P(t)-\frac{e^2}{c}\vec{A}_p(t)N_e\nonumber\\
&-&\frac{2e}{V_{cell}}\sum_{\vec{k}} \Re D^{*\vec{k}}(t)\Bigg[\vec{P}^{\vec{k}+\frac{e}{c}\vec{A}(t)}e^{iS(t)}\nonumber\\
&+& C^{\vec{k}}(t)\left(\vec{P}^{\vec{k}+\frac{e}{c}\vec{A}(t)}_{cc}+\vec{k}+\frac{e}{c}\vec{A}(t)\right)\Bigg].
\end{eqnarray}
SCM accounts for the third term of Eq.~(\ref{eq:Cur_app2}), which is the current induced by the probe pulse.

\section{Keldysh theory vs parabolic band model with Kane's matrix element} 
From the approximated expression for the $C^{\vec{k}}(t)$ (Eq.~(\ref{C_CW})), 
the total transition probability induced by the laser field, $W$, is found to be:
\begin{eqnarray}
\label{Rate_appC}
W&\approx&
 \frac{e^2 A_0^2 |P^{\vec{k}}|^2\mu^{3/2}}{2\sqrt{2}\pi } \int d\theta\sin\theta\nonumber\\
&\times&\sum_{l=l_0}^{\infty} ( J_{l-1}(\alpha,\beta)+J_{l+1}(\alpha,\beta)\Big)^2\sqrt{\zeta_l},
\end{eqnarray}
where 
$\zeta_l=l \Omega-(B_g+U_p)$,
$\theta$ is the angle between the polarization direction and  $\vec{k}$,
and $l_0$ is the maximum integer $l$ so that $\zeta_l > 0$.
The transition matrix  $|P^{\vec{k}}|^2$ is calculated using Eq.~(\ref{EQ:Kane}).
We can evaluate the reliability of our assumptions  by comparing Eq.~(\ref{Rate_appC})  with the corresponding results of the  Keldysh theory.

 The Keldysh theory assumes the band structure is,
 \begin{equation}
 \varepsilon_{c\vec{k}}-\varepsilon_{v\vec{k}}=B_g\sqrt{1+\frac{k^2}{\mu B_g}}.
 \end{equation}
 The matrix element of the optical transition from the valence to the conduction band is defined as the residue value at the saddle point of Eq.~(29) in Ref. (\cite{Keldysh65}),
 \begin{equation}
 resV_{cv}(\vec{k})=res\left[i\int u_{c}^{\vec{k}*}e\vec{E}\nabla_{\vec{k}} u_{v}^{\vec{k}}d\vec{r}\right]=\pm i\Omega/4
 \end{equation}
In contrast, we used a parabolic two-band system: i.e.
\begin{equation}
\varepsilon^G_{c,\vec{k}}-\varepsilon^G_{v,\vec{k}}=B_g+\frac{k^2}{2\mu}.
\end{equation}

 Figure \ref{fig8} shows the excitation rate for the diamond calculated using the Eq.~(\ref{Rate_appC}) (red line) and the Keldysh theory (blue line).
 The frequency of the laser is set to 1.55 eV.
The parabolic band model with Kane's transition matrix [Eq.~(\ref{Rate_appC})] shows reasonable agreement with the Keldysh theory.
This result indicates that our approach has reliability as high as Keldysh theory.
 
\begin{figure} 
\includegraphics[width=90mm]{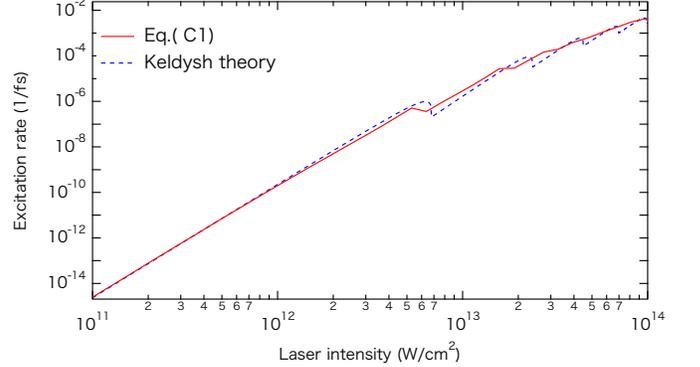} 
\caption{\label{fig9} 
Comparison of excitation rates from Keldysh theory (blue dotted line) and the calculation using Eq.~(\ref{Rate_appC}) with Kane's mode (red solid line)l.}
\end{figure}

\end{document}